\documentclass[prl, twocolumn, superscriptaddress, nofootinbib, aps, floatfix,longbibliography]{revtex4-1}
\usepackage{graphicx}
\usepackage{float}
\usepackage{subfigure}
\usepackage{dcolumn}
\usepackage{bm}
\usepackage{amsmath}
\usepackage{amsthm}
\usepackage{multirow}
\usepackage{enumerate}
\usepackage{amsfonts}
\usepackage{ifthen}
\usepackage{psfrag}
\usepackage{slashed}
\usepackage{hyperref}
\usepackage{gensymb}
\usepackage[utf8]{inputenc}
\usepackage{color}
\usepackage{subfigure}
\usepackage{ulem}
\usepackage[utf8]{inputenc}

\newcommand{\be}{\begin{equation}}
\newcommand{\ee}{\end{equation}}
\newcommand{\bea}{\begin{eqnarray}}
\newcommand{\eea}{\end{eqnarray}}

\begin{document}

\title{New Physics Opportunities in Triangle Singularity}

\author{Yu Gao}
\email{gaoyu@ihep.ac.cn}
\affiliation{Institute of High Energy Physics, Chinese Academy of Sciences, Beijing, 100049, China}

\author{Yu Jia}
\email{jiay@ihep.ac.cn}
\affiliation{Institute of High Energy Physics, Chinese Academy of Sciences, Beijing, 100049, China}
\affiliation{School of Physical Sciences, University of Chinese Academy of Sciences, Beijing, 100049, China}

\author{Yugen Lin}
\email{linyugen@ihep.ac.cn}
\affiliation{Institute of High Energy Physics, Chinese Academy of Sciences, Beijing, 100049, China}
\affiliation{School of Physical Sciences, University of Chinese Academy of Sciences, Beijing, 100049, China}

\author{Jia-Yue Zhang}
\email{zhangjiayue@ihep.ac.cn}
\affiliation{Institute of High Energy Physics, Chinese Academy of Sciences, Beijing, 100049, China}
\affiliation{School of Physical Sciences, University of Chinese Academy of Sciences, Beijing, 100049, China}

\begin{abstract}
We show that loop-induced processes involving new physics particles can readily satisfy Landau Equation and trigger triangular singularities at high energy colliders, leading to fully visible Standard Model final states. Four-particle vertices in new physics allow triangular singularity diagrams to evade large virtuality suppression. In addition, a $t$-channel triangular singularity can also occur in particle scattering processes that may extend to low momentum exchange. We discuss several typical scenarios in supersymmetric and extended Higgs models, then identify the singular component in the loop-integral amplitude at specific external momentum points.
\end{abstract}

\maketitle


{\bf Triangle Singularity} (TS)~\cite{Karplus:1958zz,Landau:1959fi}, the kinematic situation when all three intermediate particles in a triangle loop diagram for a $1\rightarrow 2$ process become on-shell, features a singularity in the scattering amplitude~\cite{Peierls:1961zz,Goebel:1964zz} that can by-pass~\cite{Aitchison:1968cho,Anisovich:1995ab,Szczepaniak:2015hya,Debastiani:2018xoi} the Schmid theorem~\cite{Schmid:1967ojm}. Since an invariant-mass resonance may masquerade as an emerging new physical state, interest in triangle singularity recently grows immensely in hadron spectroscopy as reviewed in Ref.~\cite{Guo:2017jvc,Guo:2019twa}. Famous examples include the successful explanation of a large isospin violation in $\eta(1405/1475)\rightarrow3\pi$~\cite{Aceti:2015zva,Wu:2011yx} and the $a_1(1420)$ enhancement in $\pi^- p$ scattering~\cite{Liu:2015taa,Mikhasenko:2015oxp}.

Triggering TS requires the internal momenta $k_i^\mu$ in a triangle diagram to satisfy the Landau Equation~\cite{Landau:1959fi}, 
\be
\sum_i \alpha_{i}k_i^\mu = 0 {~~\rm and ~~} k_i^2 -m_i^2=0,~~ (i=1,2,3),
\ee
where $\alpha_{i} \in [0,1]$ and should satisfy $\sum_i\alpha_i=1$. These conditions can be equivalently expressed in relation with the external invariant momenta $p^2_k$ that connect to $k_{i},k_j$, 
\bea 
\beta_i + \sum_{j}^{j\neq i} \beta_j y_{ij} &=& 0, ~\label{eq:ts2} \\
 {\rm where}~~y_{i j}&\equiv&\frac{m_{i}^{2}+m_{j}^{2}-p_{k}^{2}}{2 m_{i} m_{j}}, \nonumber
\eea
in which $i\neq j\neq k$ and $\beta_i\equiv \alpha_i m_i$. TS typically describes a heavy state splitting into lighter states with three internal on-shell particles with various masses. Thus a copious number of physical states, like that with hadrons, make it easier to satisfy singularity conditions. 

In this paper, we explore several TS scenarios with the kaleidoscopic particle content in beyond the Standard Model (BSM) theories, such as in supersymmetry and extended Higgs sector models, etc. When coupled to the Standard Model (SM), new physics (NP) particles nicely fill in the internal lines $\{k_i\}$ in a TS process with sufficiently large incident invariant momentum $\sqrt{p^2}\sim {\cal O}(1)M_{\rm NP}$ at the new physics mass scale that allows the internal NP particles to be on-shell. At a collider, fusion of two SM particles can provide this large incident invariant momentum, and the fractional parton momenta can guarantee the collision energy meets the TS's required value $\hat{s}=p^2$ when the beam is sufficiently energetic. We will discuss several characteristic new physics TS scenarios and demonstrate that the TS process can avoid large virtuality suppression with four-boson couplings, which are abundant in BSM. In addition, we will identify TS in a $t$-channel momentum exchange process that converts a massive NP state into other NP states.

{\bf Kinematics} at TS derive from Landau Equation's requirements on momenta. For clarity, we will use latin subscripts for the three external momenta $p_i$ as $\{ p_A, p_B, p_C \}$. In order for Eq.~\ref{eq:ts2} to have physical solutions, their invariant self-products $\{ p^2_A, p^2_B, p^2_C \}$ satisfy the relations ~\cite{Goebel:1964zz,Guo:2016bkl}  
\bea
p^2_C &\in &\left[\left(m_{1}+m_{2}\right)^{2}, m_{1}^{2}+m_{2}^{2}+m_{2} m_{3}+\frac{m_{2}}{m_{3}}\left(m_{1}^{2}-p_{B}^{2}\right)\right] ~ \label{eq:range}\\
p^2_A &\in &\left[\left(m_{2}+m_{3}\right)^{2}, m_{2}^{2}+m_{3}^{2}+m_{1} m_{2}+\frac{m_{2}}{m_{1}}\left(m_{3}^{2}-p_{B}^{2}\right)\right],~\nonumber 
\eea
where the internal resonant masses need to be positive and at least two internal particles have non-identical values. It is known from Eq.~\ref{eq:range} that external invariant momentum-squares $p^2_A$, $p^2_C$ must be positive and $p^2_B$ is free. This leads to two physical scenarios: 

{\it I}.~ All three invariant momentum-squares are positive, $p^2_A$, $p^2_B,$ $p^2_C>0$, which typically corresponds to a decay process or an $s$-channel collision process into two final-state momentum systems. 

{\it II}.~ One negative invariant momentum-square, say $p_B^2<0$, that can occur in a $t$-channel scattering process with $p_B$ as a {\it virtual} momentum exchange.

\begin{figure*}[t]
\subfigure[]{
\includegraphics[scale=0.4]{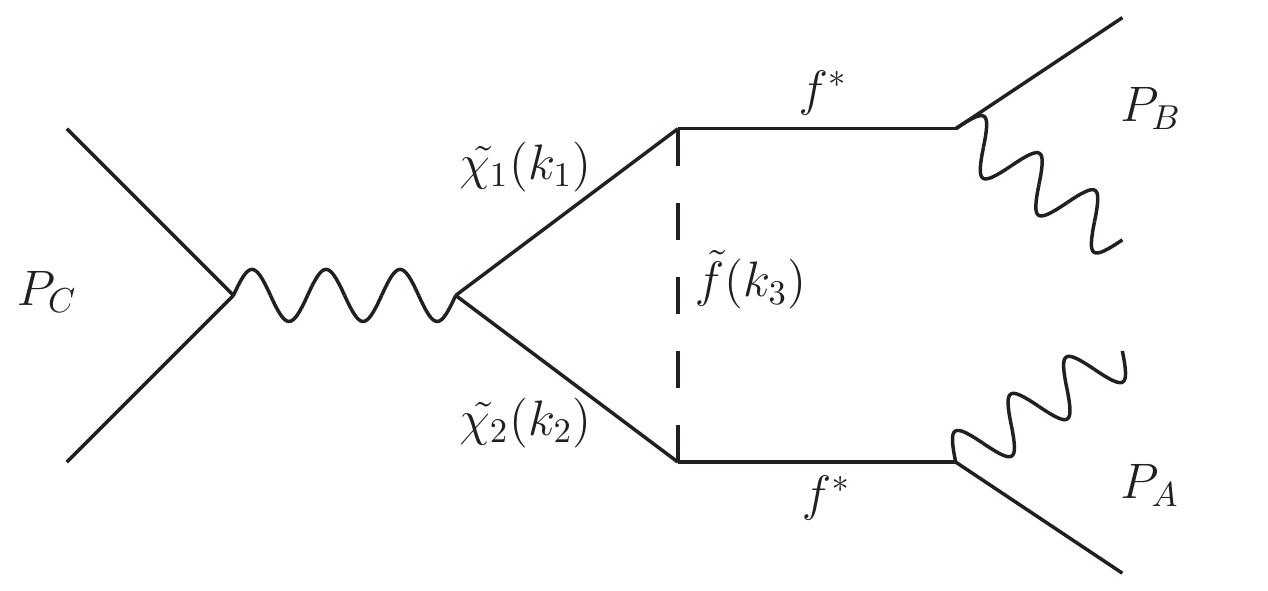}}
\subfigure[]{
\includegraphics[scale=0.4]{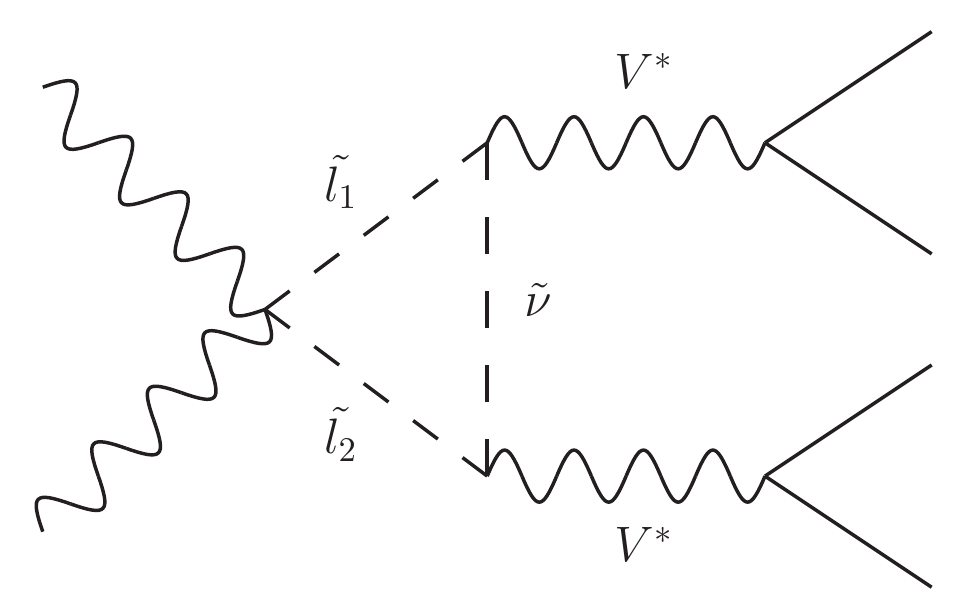}}
\subfigure[]{
\includegraphics[scale=0.4]{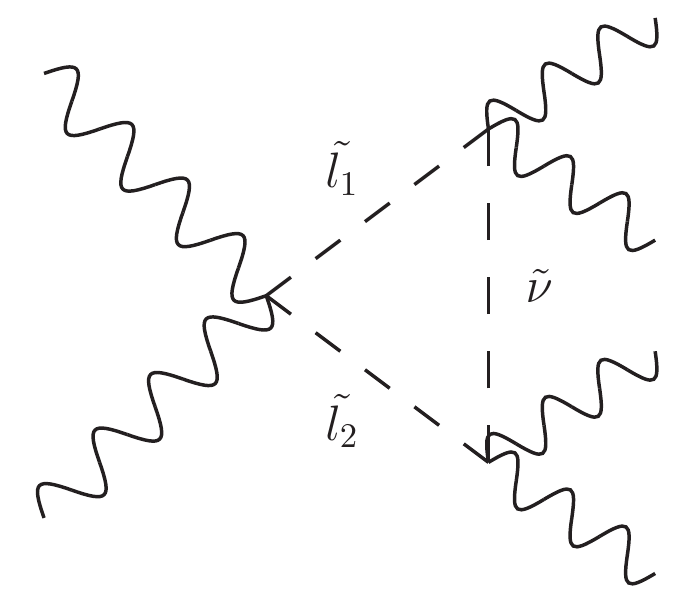}}
\subfigure[]{
\includegraphics[scale=0.35]{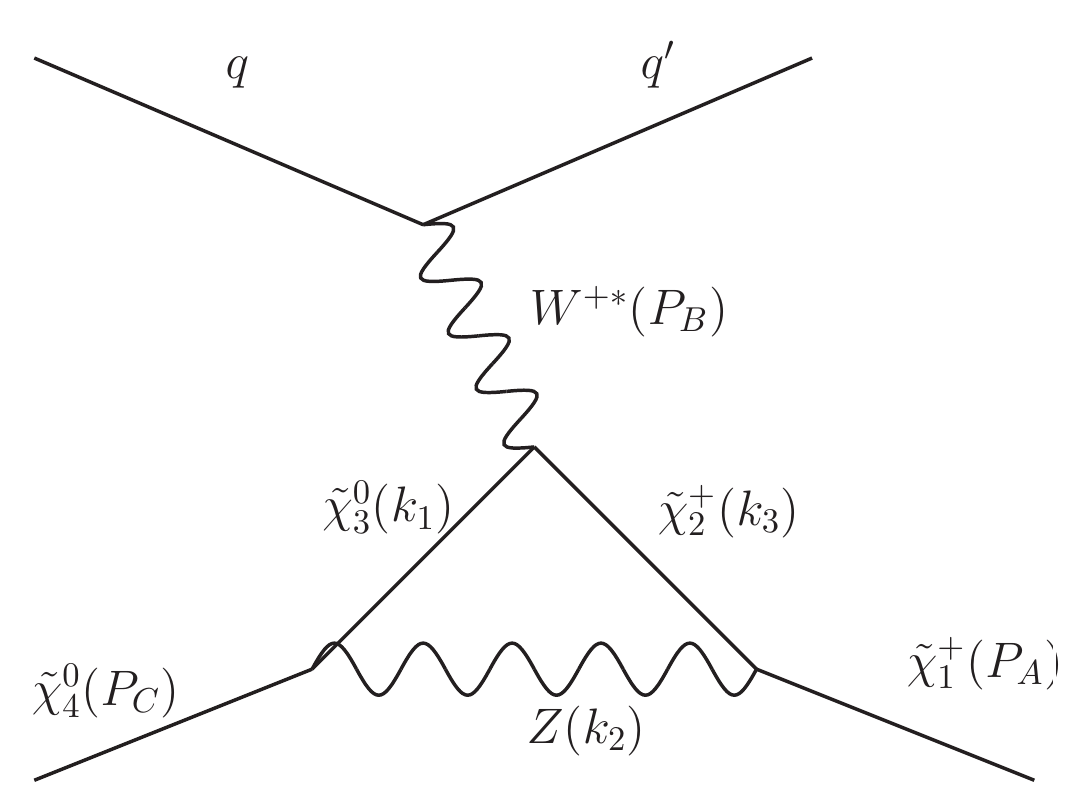}}
\caption{Characteristic triangle singularity diagrams with new physics particles in the loop, include the Drell-Yan (left) and Vector Boson Fusion (mid-left) process. Four-point vertices in a VBF diagram would allow two-particle systems to replace a virtual external line (mid-right) and avoid large propagator suppression. TS also allows one negative external $p^2$ that occurs in a scattering process with virtual momentum exchange (right). Gauginos and sleptons in the MSSM are labeled here as representatives of BSM fermions and bosons that couple to the external SM lines.}
\label{fig:process}
\end{figure*}


\medskip
In {\bf Scenario I}, fusion of two SM partons at the collider provides $\hat{s}=p_C^2$ up to the center-of-mass collision energy that dynamically covers a large range of BSM $m_1, m_2$. Typical process include the Drell-Yan and Vector Boson fusion (VBF) diagrams, as shown in Fig.~\ref{fig:process}(a), (b), (c), in which  charged BSM particles can occupy the internal line(s) $k_i$ inside the triangle loop. Taking an example with the minimal supersymmetric standard model (MSSM), a Drell-Yan process with fermionic $p_A,p_B$ realizes with neutralino/chargino(s) as $k_1,k_2$, a sfermion as $k_3$, and two virtual SM fermions as $p_A,p_B$. Alternatively an all-fermion triangle loop is possible with neutralino/charginos as $k_i$, such as with $\{\tilde\chi^-,\tilde\chi^+,\tilde\chi^0\}$, $\{\tilde\chi^0,\tilde\chi^0,\tilde\chi^-\}$, $\{\tilde\chi_i^0,\tilde\chi_j^0,\tilde\chi_k^0\}$, etc., the latter containing at least two different neutralinos. For fermion loops the $p_A,p_B$ must be bosonic, including the SM electroweak gauge bosons. Note in the SM there is also a massive $\{ t,\bar{t},b\}$ loop that can satisfy the TS condition in a $Z^*\rightarrow W^*W^*$ process. For massive loops, the $p_A,p_B$ invariant masses are generally large, $p_A,p_B$ must further split into particle pairs to become a physical final state.

In a VBF process, two initial-state radiated bosons can directly couple to $k_1,k_2$ with a four-particle vertex, in addition to the diagram of two boson first fusing into an $s$-channel propagator. Renormalizable four-particle couplings would require BSM bosons as $k_i$. In the MSSM, possible examples include slepton/squarks completing the triangle loop, e.g. with $\{ \tilde{l},\tilde{l},\tilde{\nu} \}, \{ \tilde{q},\tilde{q},\tilde{q}' \}$, where the two $\tilde{l}$ can either be the same slepton, or a combination of two different slepton mass eigenstates $\tilde{l}_{1},\tilde{l}_{2}$ after the mixing of left/right-handed sfermions in the MSSM, and so in the case of the squarks. Here the MSSM sfermion triangle serve as one example; well-motivated candidates also include extra scalar fields in extended electroweak models~\cite{Gunion:1989we}, leptoquarks in GUT models~\cite{Georgi:1974sy,Pati:1974yy} and numerous other exotics will fit into the diagram.

Note triggering TS requires different internal masses, and this requires a BSM theory to possess enough number of extra particles to fill in the loop. For instance, in `nHDM', or extended weak scalar-sector models with $n\ge 2$ copies of Higgs doublets, mixing of these Higgs fields yield $n$ neutral $h^0$ scalars and $n-1$ charged $h^+$ scalars after electroweak symmetry breaking. So that an $\{h^0,h^0,h^+\}$ triangle that connects $ZW^+W^-$ is present in 2HDM, while a triangle that connects $\gamma ZZ$ requires two different charged scalars, $\{h_i^+,h_i^+,h_j^+\}$ with $i\neq j$, namely asking for another charged scalar in the model, like in $(n\ge 3)$HDM. 

A complication rises from GIM theorem~\cite{Glashow:1970gm} that the couplings to gauge bosons should vanish if two different particles carry exactly the same charge. As a result, pure photonic coupling is only allowed for two internal particles with the same electric charge but different masses, and the TS triangle can not connect to more than one external purely-photon line. Coupling with other weak bosons are generally not restricted by GIM as BSM particles can have mass origins other than one single Higgs vev. In our showcase diagram with same-charge particles, e.g. MSSM slepton/sneutrino loop with a mixture of $\tilde{l}_{1}, \tilde{l}_{2}$, or nHDM loop with different charged scalars, at least one $W$ or $Z$ boson must be present on such a vertex.

Internal lines being on-shell requires $\sqrt{p_C^2}$ to be higher than the sum of $m_1$ and $m_2$ at TS. In case of a BSM triangle loop, the center-of-mass energy $\sqrt{\hat s}>m_1+m_2$ means TS only triggers (slightly) above BSM production threshold, namely TS is still a {\it near-threshold} phenomenon at the BSM scale. Yet TS differs from BSM production in both the narrow resonant lineshape and a very different final state: TS can produce a final state purely composed of visible SM particle-systems that carry BSM scale energies. This would help reveal the so-called {\it `compressed'} BSM particle spectrum at the collider, in which heavy BSM particle masses lay very close to that of an invisible particle.

\medskip
In {\bf Scenario  II}, we consider a negative external $p_B^2$, which corresponds to a $t$-channel momentum exchange from the external environment. Physically, a negative $p_B^2$ prevents a spontaneous $C\rightarrow A+B$ decay as $p^2$ for a sum of physical momenta is always non-negative. Instead, it allows a massive system $C$ to {\it exchange momentum} during a collision process and converts to particle system $A$, as illustrated in the last diagram in Fig.~\ref{fig:process}. 
For this case, we choose $m_3>m_1$ so that spontaneous decay would not occur and particle 1 must receive external momentum to realize $1+B \rightarrow 3$ process. Besides, since $C$ splits into on-shell $m_1,m_2$ at TS, $C$ can not be the lightest stable state of a decay-able particle spectrum (like a LSP dark matter), and the $t$-channel process refers to an incited conversion with momentum transfer from the environment. In Fig.~\ref{fig:process}(d) we illustrate with the conversion of an MSSM neutralino into one chargino through a $\{\tilde\chi^0,\tilde\chi^+,Z\}$ triangle loop that exchanges a virtual $W$ boson with an external charged current. Such processes will apply to meta-stable BSM particles when they were still present in the (early) Universe. Note in the $t$-channel conversion $C$ is usually lighter than the invariant mass of final-state system $A$, due to the in-flow of external momentum $p_B$.

For high-energy BSM search at the collider, forming a massive $p_C^2>0$ can be difficult as a typical collider $e,p$ beam does not have {\it on-shell} heavy partons. Fusion of two initial-state partons into $p_C^2$ would complicate the whole scattering into a three body process. Interestingly, 
a negative $p_B^2$ can approach to zero in the limit 
\begin{equation}
\left\{\begin{array}{l}
m_1 \rightarrow m_3 \\
p_C^2 \rightarrow p_A^2 
\end{array}\right. {\rm{~for ~}} p_B^2 \rightarrow 0.
\end{equation}
In case $|\vec{p}_B|\ll \sqrt{p_C^2},\sqrt{p_A^2}$, it would correspond to relatively soft scattering and indicates for a TS possibility in the low-energy scattering with massive/compound $C,A$ systems, which can be pursued in future research.

\medskip
{\bf Four-particle vertices} can play a special role in high energy TS diagrams when the external lines are required to be very massive. At a large collision energy, $\sqrt{p_A^2}, \sqrt{p^2_C}$ can be much higher than the weak scale ($\sqrt{p_B^2}$ may be or not which depends on the difference between $m_1$ and $m_3$), and identifying the external line with one (intermediate) single SM particle, as shown the first two diagrams in Fig.~\ref{fig:process}, will suffer at least two major propagator virtuality suppression $\propto (p_{A,C}^2 -m_{\rm SM}^2)^{-1}\sim p_{A,C}^{-2}$ in the scattering amplitude. Thus for a large $p^2$ it is kinematically favorable to replace $p$ with two external momenta via a four-particle coupling, such as two VBF $W^*$ bosons directly couple to the triangle vertex, and similarly, even with upto all three triangle vertices as four-vertex couples that leads to four final-state on-shell SM scalar/vector bosons, as shown in the 3rd diagram in Fig.~\ref{fig:process}. Viable four-point vertices preferably couple to two lighter ($\sum m <\sqrt{p^2}$) external particles, so that they emerge on-shell.  For heavy BSM particles above the weak scale, vertices with di-vector-boson and di-Higgs couplings make excellent options as the final state on-shell bosons can be experimentally identified.

\begin{figure}[b]
\includegraphics[scale=0.6]{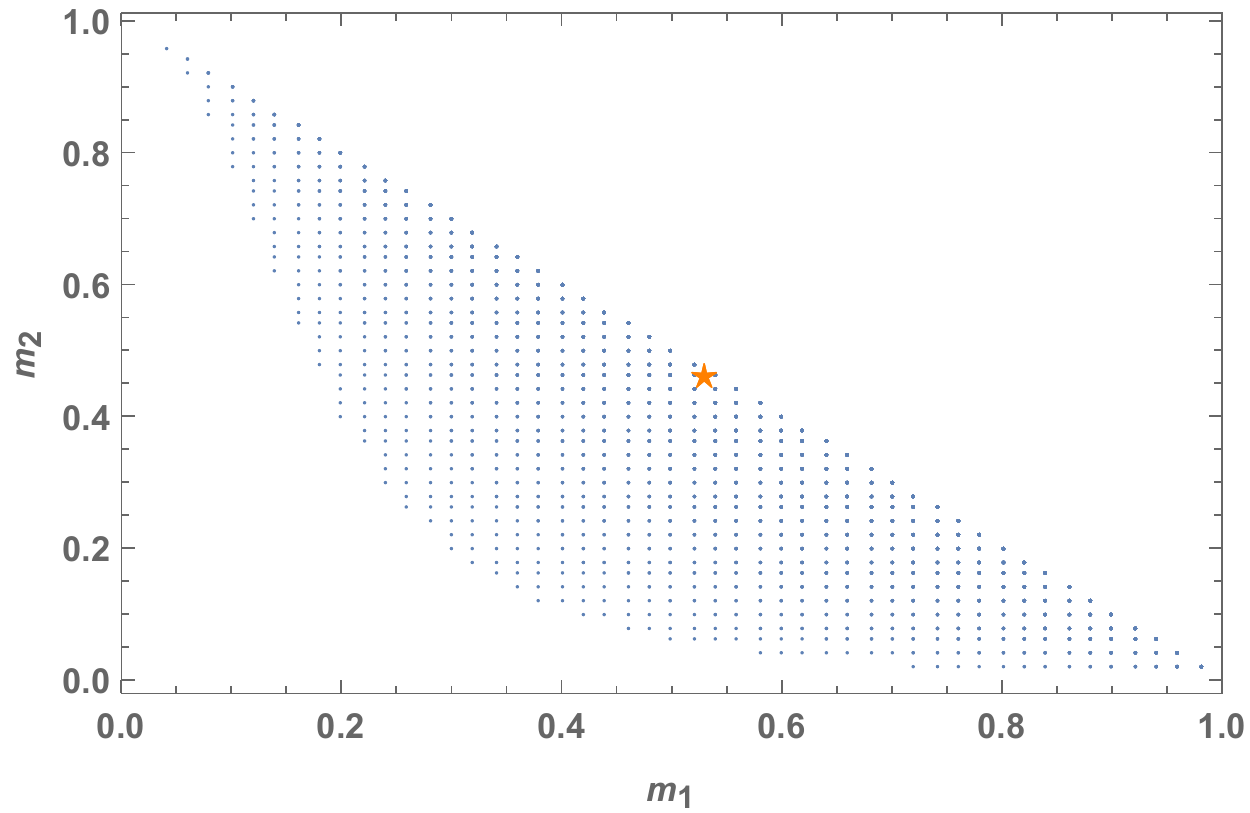}
\includegraphics[scale=0.6]{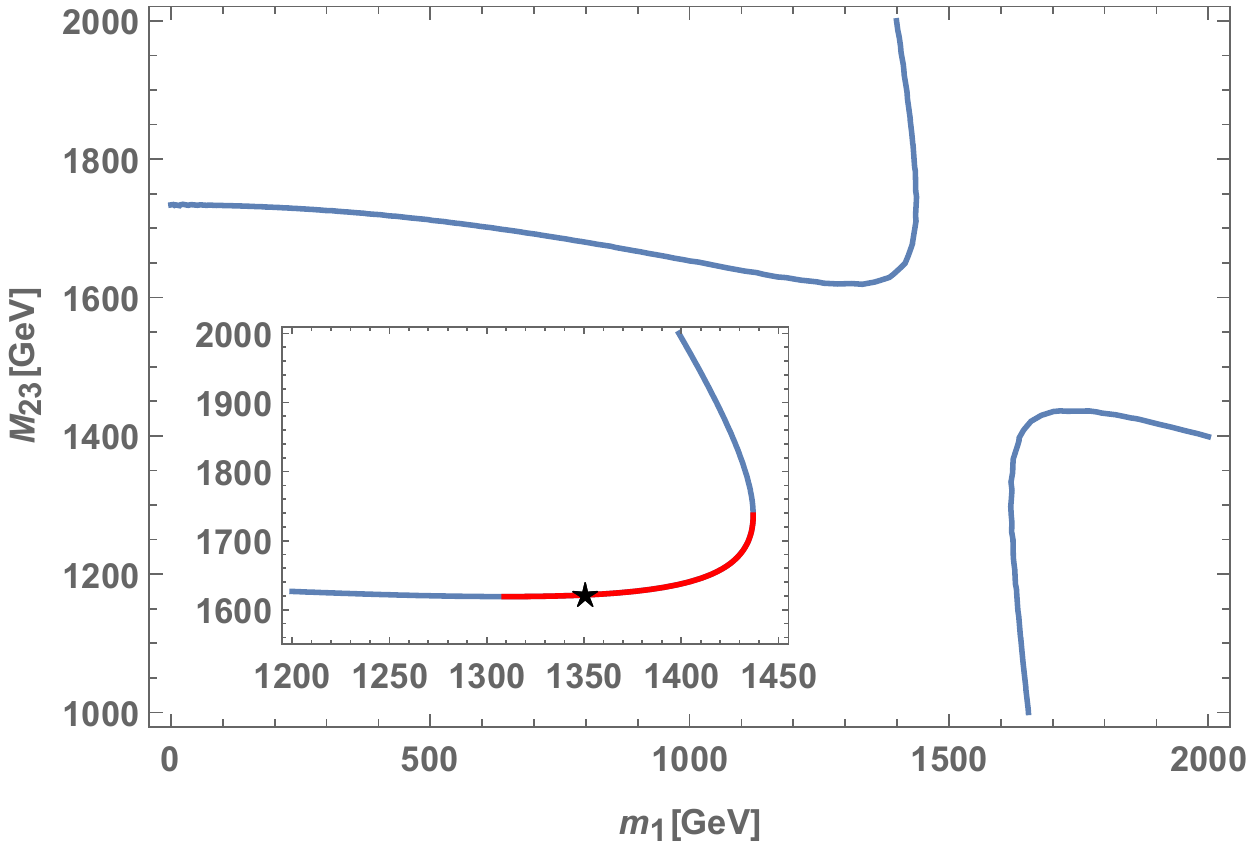}
\caption{BSM particle mass ranges (upper) that can trigger TS at a collider center-of-mass energy $\sqrt{p_C^2}$ and particle mass is normalized to $\sqrt{p_C^2}$. Scenario $I$ benchmark parameter point for WW fusion with $m_{1,2,3}=\{1701, 1500, 1499\}$ GeV at given $\sqrt{p_C^2}=3202$GeV is marked with an asterisk symbol. The lower panel shows the Dalitz plot for Scenario {\it II}, where the blue curves represent the solution of Landau Equation and the red region is the physical boundary at TS. The benchmark parameter point with \{$m_1=1351$~GeV, $M_{23}=1621$~GeV\} is also marked with asterisk symbol.}
\label{fig:range_dalitz}
\end{figure}

\medskip
{\bf Collider reach} for TS can be calculated by BSM particle masses that satisfy the conditions in Eq.~\ref{eq:ts2}. In terms of $m_1$ and $m_2$, the valid parameter space for the given collision center-of-mass energy $\sqrt{p_C^2}$ is shown by the shaded area in the upper panel of Fig.~\ref{fig:range_dalitz} where Eq.~\ref{eq:ts2} can possess physical solutions and particle mass is normalized to $\sqrt{p_C^2}$. The upper boundary of the shaded region corresponds to Eq.~\ref{eq:range} lower boundary, i.e. $m_1 +m_2 \le \sqrt{p_C^2} \le \sqrt{\hat s}$, and the lower boundary of the shaded region corresponds to the Eq.~\ref{eq:range} upper boundary. In case internal lines 1 and 2 assume the same particle, i.e. $m_1=m_2$, the maximal and minimal masses are respectively $\sqrt{p_C^2}$/2 and $\sqrt{p_C^2}$/4. Since the beam particle's parton distribution allows $\sqrt{p_C^2}$ to be anywhere below the maximal collision energy $\sqrt{\hat s}$, the subspace where ${m_1,m_2}$ are too low to hit TS at a given $\sqrt{p_C^2}$ can still trigger TS at a lower value. This allows a high energy collider to sweep through the relevant TS parameter space in an efficient way similar to that for $m_1,m_2$ pair-production.

\begin{table*}[t]
    \centering
    \begin{tabular}{c|c|ccc|ccc}
    \hline \hline
    Loop particles& Process & $m_1(\Gamma_1)$ (GeV) & $m_2(\Gamma_2)$ (GeV) & $m_3(\Gamma_3)$ (GeV) & $p_A^2$ (GeV$^2$)& $p_B^2$ (GeV$^2$) & $p_C^2$ (GeV$^2$) \\
    \hline
    $\{\tilde\chi^-_2,\tilde\chi^+_1,\tilde\chi^0_1\}$ & DY-like & 1528(13) & 359(0.9) & 306(0) & 700$^2$ & 800$^2$ &1943$^2$  \\
     $\{\tilde{l_1},\tilde{l_2},\tilde{\nu}\}$ & WW fusion & 1701(1.56) & 1500(0.34) & 1499(0.16) & 3000$^2$ & 100$^2$ & 3202$^2$ \\
   $\{\tilde\chi^0_3,Z,\tilde\chi^+_2\}$ & $t$-channel & 1351(3.79) & 91(2.45) & 1528(13) & 1621$^2$ & -2500$^2$& 1528$^2$ \\
   \hline \hline
    \end{tabular}
    \caption{Sample TS external momentum points for the process in  Fig.~\ref{fig:process}a, ~\ref{fig:process}b and ~\ref{fig:process}d, also with internal particle mass and width (in bracket). One external momentum-square is free and here we list the numerical solution for one chosen value.}
    \label{tab:masses}
\end{table*}

Next we will demonstrate the relation of internal $m_i$ and external $p^2$ at TS. In order for Eq.~\ref{eq:ts2} to have solutions, the determinant of $\beta_i$'s coefficients should be zero:
\begin{equation}
\left|\begin{array}{ccc}
1 & y_{12} & y_{13} \\
y_{12} & 1 & y_{23} \\
y_{13} & y_{23} & 1
\end{array}\right|=1+2 y_{12} y_{23} y_{13}-y_{12}^{2}-y_{23}^{2}-y_{13}^{2}=0,
\label{eq:dalitz}
\end{equation}
which contain 6 kinematic parameters. The common practice is to use a Dalitz plot to illustrate their relations, as shown in the lower panel of Fig.~\ref{fig:range_dalitz}. For variables, we follow the conventional choice of the internal mass $m_1$ and the invariant mass of 2+3 system $M_{23}=\sqrt{p_A^2}$, and the rest parameters, i.e. the internal mass $m_2$, $m_3$ and external $p_B^2$, $p_C^2$, are fixed at given values as listed in Table~\ref{tab:masses}. The corresponding trajectory of $m_{1}$ and $M_{23}$ is shown as the blue curve in the lower panel of Fig.~\ref{fig:range_dalitz}, in which the red section is the `physical boundary' which represent the physical solution region that satisfies Eq.~\ref{eq:range}. The left and right ends of the red section corresponds to the minimal and maximal $m_1$ at given \{$p_C^2$, $p_B^2$, $m_2$, $m_3$\} values. 
As the Dalitz pattern for an $s$-channel process is well known, see Ref.~\cite{Guo:2019twa} for good examples. Here, we only show the Dalitz plot for the $t$-channel process which has a different shape, and we mark our benchmark point as an asterisk symbol in the physical region.

\medskip

Next we will adopt several benchmark scenarios for TS calculation. The Drell-Yan like, VBF and $t$-channel diagrams shown in Fig.~\ref{fig:process} can be realized in MSSM with masses listed in Table~\ref{tab:masses}. When internal line masses are fixed, we can choose one external momentum-square freely, then the rest of external momentum-squares for triggering TS can be completely solved by combining Eq.~\ref{eq:ts2} and Eq.~\ref{eq:range}. In Table~\ref{tab:masses}, the involved MSSM particle masses are freely chosen in the currently unexcluded range~\cite{Sarkar:2021lju,CMS:2021cox}, and one $p_C^2$ ($p_B^2$) value is chosen for the $s$($t$)-channel scenario. 

The benchmark MSSM point's parameters are listed below: $\tan\beta$=10, \{$\mu$, $M_1$, $M_2$, $Ml_3$, $Mr_3$\}=\{300, 1500, 1400, 1500, 1700\} GeV. In our $s$-channel cases, MSSM particles can only appear inside the triangle; in the $t$-channel case, however, the external line $p_C$ and $p_A$ need to be massive, so we assign them a neutralino $\tilde{\chi}^0_4$ and a chargino $\tilde{\chi}^+_1$. Note that the initial state $\tilde{\chi}^0_4$ is on its mass shell and the final system $\tilde{\chi}^+_1$ is likely off-shell ($p^2>m^2$) which will further convert to other NP and SM states. The MSSM mass spectrum, couplings and other model parameters are computed under the `electroweak symmetry breaking' scheme (EWSB-MSSM)~\cite{Batra:2008rc} in the SUSPECT2 package~\cite{Djouadi:2002ze}. Due to the large number of model parameters, only the most relevant ones are listed here; other MSSM masses are left to mutli-TeV or their default values. 

\begin{figure}[h]
\subfigure[~$WW$ fusion]{
\includegraphics[width=.4\textwidth]{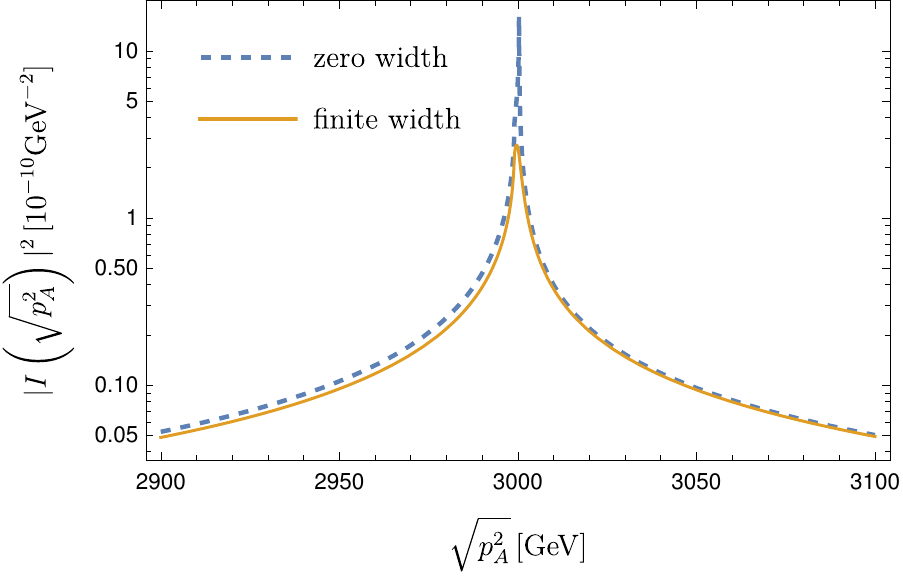}
}
\subfigure[~$t$-channel]{
\includegraphics[width=.4\textwidth]{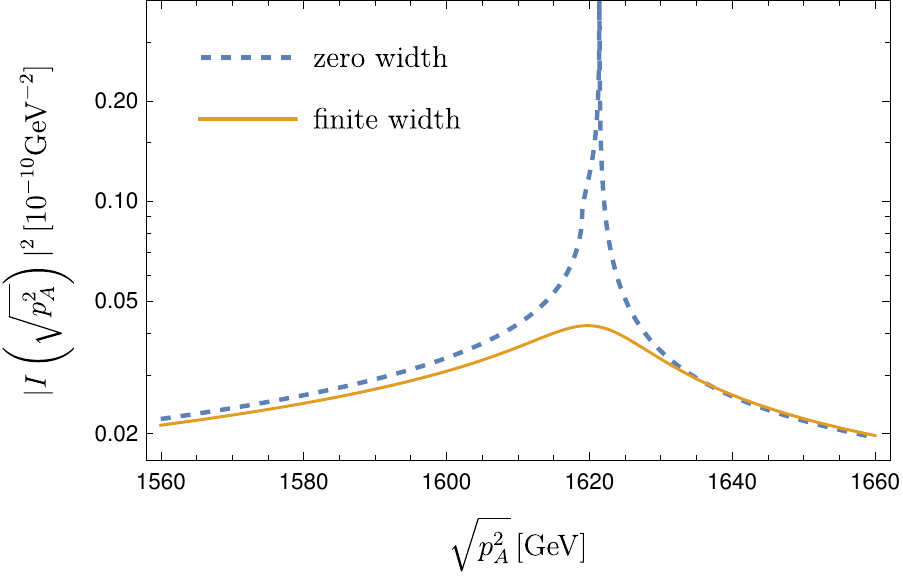}}
\caption{Modulus-square of the scalar integral in Eq.~\ref{eq:C0} with internal particles illustrated in Fig~\ref{fig:process}b (upper panel) and Fig.~\ref{fig:process}d (lower panel). The MSSM masses and widths correspond to Table.~\ref{tab:masses}.}
\label{fig:peak}
\end{figure}

\medskip
TS features an external momenta dependence in the loop amplitude that becomes singular when Eq.~\ref{eq:ts2} and Eq.~\ref{eq:range} are met. A general triangular loop amplitude can be expressed as the linear combinations of $1$-, $2$-, $3$-point scalar integrals via the standard Passarino-Veltman reduction~\cite{Passarino:1978jh}. 
The TS effect is fully-encoded in the following $3$-point scalar integral
\begin{equation}
\begin{split}
I\left(\sqrt{p_A^2}\right)=\int \frac{\mathrm{d}^{4} l}{i \pi^{2}}&\left[\frac{1}{l^2-m_3^2}\cdot \frac{1}{(l+p_A)^2-m_2^2}\right.\\
&\left. \cdot \frac{1}{(l+p_A+p_C)^2-m_1^2}\right].
\end{split}
\label{eq:C0}
\end{equation}

When the masses of internal particles and $p_B^2$, $p_C^2$ are within the range for the TS to be in the physical region, there would be a logarithmic divergence when $p_A^2$ satisfies the Landau equation. In practice, the internal particles' width will provide a small imaginary part to the propagators by replacing $m_i^2$ by $m_i^2-i m_i\Gamma_i$ in Eq.~(\ref{eq:C0}). Therefore the singularity collapses to a finite peak that broadens with the particle widths. In Fig.~\ref{fig:peak}, we plot the modulus square of integral $I\left(\sqrt{p_A^2}\right)$ for the processes shown in Fig.~\ref{fig:process}(b) and Fig.~\ref{fig:process}(d).

\medskip
{\bf To summarize,} the diverse particle spectrum in BSM theories can provide candidate particles to fill in a triangle loop diagram and satisfy triangular singularity at a high energy collider. Although TS is a near-threshold phenomenon, its manifestation with BSM loops can lead to a fully identifiable SM final state, offering a unique opportunity to search for new physics.  We discussed several specific cases in MSSM and extended Higgs models that satisfy TS with Drell-Yan and vector boson fusion processes at collider, where BSM four-point vertices can play a role of evading large virtuality suppression. Besides, we showed that $t$-channel scattering also triggers TS with virtual momentum exchange, different from traditional $s$-channel decay processes. For the loop integral, we identified and illustrated the singular amplitude component at benchmark $s$($t$)-channel TS scenarios.

\medskip
{\bf Acknowledgements.}\\
Y.~G. and Y.~Lin thank for support by the Institute of High Energy Physics, Chinese Academy of Sciences (E2545AU210) and in part by the National Natural Science Foundation of China (12150010). The work of Y.~J. and J.-Y.~Z. is supported in part by the National Natural Science Foundation of China under Grants No. 11925506, 11875263, No. 11621131001 (CRC110 by DFG and NSFC).

\bibliography{refs}

\end{document}